\documentclass[reprint,showpacs,pra]{revtex4}
\usepackage[T1]{fontenc}
\usepackage[latin9]{inputenc}
\usepackage{color}
\usepackage{amsmath}
\usepackage{amssymb}
\usepackage{graphicx}
\usepackage{esint}
\usepackage{bm}
\usepackage[english]{babel}
\usepackage{appendix}

\newcommand{\diag}{\operatorname{diag}}
\makeatletter \@ifundefined{textcolor}{} {
\definecolor{BLACK}{gray}{0}
 \definecolor{WHITE}{gray}{1}
 \definecolor{RED}{rgb}{1,0,0}
 \definecolor{GREEN}{rgb}{0,1,0}
 \definecolor{BLUE}{rgb}{0,0,1}
 \definecolor{CYAN}{cmyk}{1,0,0,0}
 \definecolor{MAGENTA}{cmyk}{0,1,0,0}
 \definecolor{YELLOW}{cmyk}{0,0,1,0}
 }
\begin{document}

\title{Polarimetric measurements of single-photon geometric phases}
\author{O. Ort\'{i}z$^{1}$, Y. Yugra$^{1}$, A. Rosario$^{1}$, J. C. Sihuincha$^{1}$, J. C. Loredo$^{2}$, M. V. Andr\'{e}s$^{3}$ and F. De
Zela$^{1}$}

\affiliation{$^{1}$ Departamento de Ciencias, Secci\'{o}n
F\'{i}sica, Pontificia Universidad Cat\'{o}lica del Per\'{u}, Apartado 1761, Lima, Peru.%
\\
$^{2}$ Centre for Engineered Quantum Systems, Centre for Quantum Computer and Communication Technology,and School of Mathematics and Physics, University of Queensland, 4072 Brisbane, QLD, Australia\\
$^{3}$  Departamento de F\'{i}sica Aplicada y Electromagnetismo, Universidad de Valencia, c/Dr. Moliner 50, Burjassot, Valencia, Spain}

\begin{abstract}
We report polarimetric measurements of geometric phases that are
generated by evolving polarized photons along non-geodesic
trajectories on the Poincar\'{e} sphere. The core of our
polarimetric array consists of seven wave plates that are
traversed by a single photon beam. With this array any $SU(2)$
transformation can be realized. By exploiting the gauge invariance
of geometric phases under $U(1)$ local transformations, we nullify
the dynamical contribution to the total phase, thereby making the
latter coincide with the geometric phase. We demonstrate our
arrangement to be insensitive to various sources of noise entering
it. This makes the single-beam, polarimetric array a promising,
versatile tool for testing robustness of geometric phases against
noise.
\end{abstract}

\pacs{03.65.Vf, 03.67.Lx, 42.65.Lm}
\maketitle

\section{Introduction}

Even though experiments testing different properties of geometric
phases are continuously reported, theoretical developments can
expand at such a spanking pace that experimental testing can be
left behind for a while. This seems to be the case with the
subject of geometric phases. Since Berry's seminal work
\cite{berry}, which brought to light the appearance of geometric
phases in adiabatically evolving, cyclic quantum processes, there
have been considerable generalizations of the subject. From Hannay
angles in the classical domain \cite{hannay} to geometric phases
in mixed quantum states subjected to non-unitary and non-cyclic
evolutions \cite{tong,peixoto,cucchietti,uhlmann,sjoqvist}, the
original concept of geometric phases has been widely expanded.
Experimental testing is required not only because of fundamental
reasons lying at the basis of all empirical sciences, but because
experimental input can help us in finding the answer to open
questions. Notably, the question about a proper, self-consistent
definition of a geometric phase for non-unitary evolutions still
remains open \cite{marzlin,bassi,rezakhani,yin,buric,lombardo}.
Similarly, the kind of robustness that geometric phases might have
against decohering mechanisms is also an open question of utmost
importance, particularly in the realm of quantum computation
\cite{sjoqvist2}. It is thus useful to explore as much
experimental techniques as possible. One should not refrain from
mirroring experiments already performed with one technique and
conduct similar experiments based on another independent
technique. This can provide not only new insights, but an enlarged
versatility as well. Geometric phases are particularly well suited
for such an approach, as they notoriously appear in the evolution
of two-level systems. Such systems can be realized under manifold
situations, quantal and classical ones. The drawbacks of one
technique could then be replaced by some advantages of the other.
For example, the physical realization of the qubit as a spin
one-half particle, e.g. a neutron, has its counterpart in the
realization of the qubit as a polarized photon. While as a source
of the former one needs a nuclear reactor, as a source of the
latter it suffices a diode-laser. On the other hand, the
versatility reached in experiments with neutrons can outperform
the one reached with their optical counterparts. A challenge is
thereby put on the latter, as to how to improve their versatility.
We have addressed such a challenge in the present work. We report
on experiments performed with single photons, which to some extent
mirror previous experiments that were conducted with neutrons
\cite{klepp,bertlmann,klepp2,filipp}. Our experiments put under
test theoretical predictions about $SU(2)$ evolutions along
non-geodesic paths. Using neutrons, experiments along these lines
have been conducted by exploiting the advantages offered by
polarimetric techniques. In contrast to interferometric techniques
\cite{werner}, polarimetric ones have an intrinsic robustness,
because they require a single beam \cite{wagh1}. The challenge
posed here, however, is how to manipulate two coherently
superposed states that are not spatially separated. In
interferometry, the (binary) path degree of freedom can be used
together with an ``internal'' degree of freedom, e.g. the spin,
that is carried along by the particle. In polarimetry instead,
there is only one path. One must then figure out how to deal with
this restriction and nevertheless reach a versatility that is
comparable to that of interferometry. The latter offers, for
example, the possibility of spin-path entanglement. In neutron
polarimetry, energy-polarization and even a tripartite
energy-polarization-momentum entanglement have been achieved
\cite{sponar}. Although an all-optical version of the latter seems
difficult to implement, there are other features that can be
exploited with advantage in optical polarimetry. We show here how
to exploit the invariance of geometric phases under local gauge
transformations \cite{mukunda}, in order to nullify the dynamical
part of the total (Pancharatnam) phase \cite{pancharatnam},
thereby making this phase coincide with the geometric phase. What is meant by gauge invariance is the invariance under the change $%
|\psi (s)\rangle \rightarrow |\psi ^{\prime }(s)\rangle =\exp
\left( i\alpha (s)\right) |\psi (s)\rangle $ of an unitarily
evolving state $|\psi (s)\rangle $. By exploiting this invariance,
one can nullify the dynamical contribution to the total phase
$\Phi _{P}=\arg \langle \psi (s_{1})|\psi (s_{2})\rangle $ between an initial and a final state, $|\psi
(s_{1})\rangle $ and $|\psi (s_{2})\rangle $, respectively. What
remains after elimination of the dynamical part is the purely
geometric contribution $\Phi _{g}$ to the total phase $\Phi
_{P}=\Phi _{g}+\Phi _{dyn}$. The $SU(2)$ evolutions we have
addressed are those of the type given by $U_{n}(\theta ,\varphi
,s)=\exp \left[ -is\mathbf{n}(\theta ,\varphi )\cdot \boldsymbol{\sigma }/2%
\right] $. Here, $\mathbf{n}$ is a unit vector,
$\boldsymbol{\sigma }$ is the triple of Pauli matrices and $s$ is
the rotation angle (on the Bloch or Poincar\'{e} sphere). We could
generalize our approach so as to deal with unit vectors that
depend on $s$, but we have focused on cases with a fixed
$\mathbf{n}$. We also restricted ourselves to deal with pure
single-photon states. These restrictions are justified in view of
the extension already achieved by considering the production of
geometric phases in systems subjected to transformations
$U_{n}(\theta ,\varphi ,s)$ of the above type. Previous
experimental tests were restricted to particular trajectories that
a system follows when subjected to some special transformations
\cite{klepp,klepp2,filipp}. The cases we address here let us study
what happens when we lift these restrictions. In such a case, a
series of features shows up that is worthwhile to analyze before
undertaking a systematic investigation of, say, the sensitivity of
geometric phases to environmental influences. A main motivation of
the present work was to analyze and to explain the appearance of
the aforementioned features. This opens the way for using this
array as a basic component for testing the impact of decohering
mechanisms.

\section{Polarimetry}

The standard procedure to exhibit the relative phase between two
states is to make them interfere and then record the intensity of
the interfering pattern by varying the relative phase. An
archetypical setup for doing this is a Mach-Zehnder
interferometer. Expressed in the language of quantum gates
\cite{Audretsch}, such a device consists of two Hadamard gates --
i.e., two beam splitters -- and a phase-shifter. A Hadamard gate
can be represented in terms of Pauli matrices as $U_{H}=(\sigma
_{x}+\sigma _{z})/\sqrt{2}$ , while the phase-shifter can be
represented as $U_{\phi }=\exp (-i\phi \sigma _{z}/2)$.
Hereby, we establish a one-to-one correspondence between the eigenvectors $%
\left\vert \pm \right\rangle $ of $\sigma _{z}$ and the two paths of the
interferometer. The action of the interferometer on an input state $%
\left\vert +\right\rangle $ is thus given by $\left\vert
+\right\rangle \rightarrow U_{H}U_{\phi }U_{H}\left\vert
+\right\rangle $. The output intensity that is recorded at, say, a
$\left\vert +\right\rangle $-detector, reads $I=\left\vert
\left\langle +\right\vert U_{H}U_{\phi }U_{H}\left\vert
+\right\rangle \right\vert ^{2}=(1+\cos \phi )/2$. Now, instead of
assigning the states $\left\vert \pm \right\rangle $ to the two
possible paths of the interferometer, we can make them correspond
to the horizontal and vertical polarization states of a single
light-beam. We thereby change from
interferometry to polarimetry. In the latter, the action of $U_{\phi }$ and $%
U_{H}$ can be realized with the help of quarter-wave ($Q$) and
half-wave ($H$) plates. Indeed, we have that $U_{\phi }=Q(\pi
/4)H((\phi -\pi )/4)Q(\pi /4)$ and $U_{H}=-iH(\pi /8)$. The
arguments in $H$ and $Q$ refer to the angles made by the plate's
major axis and the vertical direction. Up to a global phase, the
action of the Mach-Zehnder interferometer can then be mirrored in
polarization space by letting a polarized light-beam traverse a
gadget that consists of a couple of aligned retarders. In the
present case, such an array is given by $Q(\pi /2)H((2\pi -\phi
)/4)Q(\pi /2)$. This last expression is obtained by using
$Q(\alpha )H(\beta )=H(\beta )Q(2\beta -\alpha )$ and $Q(\alpha
)H(\beta )H(\gamma )=Q(\alpha +\pi /2)H(\alpha -\beta +\gamma -\pi
/2)$. Hence, by setting a horizontal polarizer before a detector
and recording the intensity as a function of $\phi $, we get a
pattern that looks the same as the interferogram produced with the
Mach-Zehnder device. Polarimetry has the great advantage of being
largely insensitive to those perturbations that in the case of
interferometry lead to random phase shifts. On the other hand, the
states $\left\vert \pm \right\rangle$ cannot be individually
addressed, as they are no longer spatially separated from one
another, as it occurs in interferometry. We must then find a way
to extract the desired information by adequately projecting the
manipulated states before detection. In the case of geometric
phases this is indeed possible, as we show next.

\begin{figure}[tbp]
\begin{centering}
\includegraphics[scale=0.55]{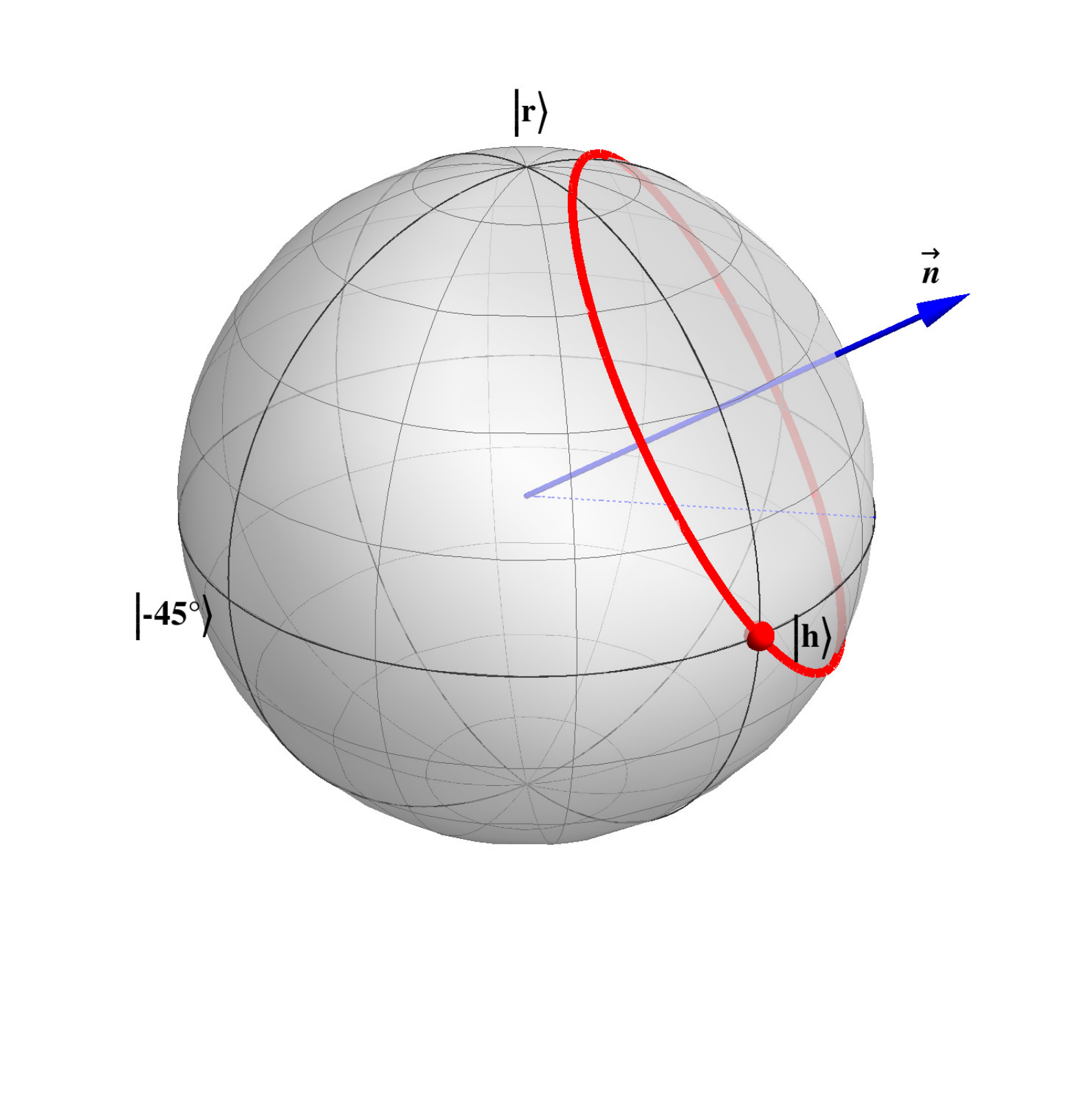} %\includegraphics[angle=-90,scale=1.0]{multi.eps}
\end{centering}
\caption{(Color online) Path followed on the Poincar\'{e} sphere by the Stokes vector that corresponds to an initial state $|h\rangle$ being submitted to a transformation $\exp(-is\mathbf{n} \cdot \boldsymbol{\sigma}/2)$. The rotation axis $\mathbf{n}$ has polar angles $\theta=\pi/3,\varphi=\pi/4$. The dynamical contribution to the total phase $\Phi_{P}$ is gauged-away all along the curve, so that $\Phi_{P}=\Phi_{g}$ holds at each value of $s$.}\label{path}
\end{figure}

Following a similar procedure as the one introduced by Wagh and
Rakhecha \cite{wagh1} -- thereby extending to single photons some
techniques already employed with classical light
\cite{loredo,fdz,loredo2} -- we consider an initial, horizontally
polarized state $\left\vert h\right\rangle $ and submit it to a
$\pi /2$-rotation around the $x$-axis. This produces a circularly
polarized state $\left( \left\vert h\right\rangle -i\left\vert
v\right\rangle \right) /\sqrt{2}$. By submitting this state to the
transformation $\exp \left( -i\phi \sigma _{z}/2\right) $ we get
$V\left\vert h\right\rangle \equiv \exp \left( -i\phi \sigma
_{z}/2\right) \exp \left(
-i\pi \sigma _{x}/4\right) \left\vert h\right\rangle $, which is the state $%
\left( \left\vert h\right\rangle -ie^{i\phi }\left\vert v\right\rangle
\right) /\sqrt{2}$, up to a global phase. Hence, we have generated a
relative phase-shift $\phi -\pi /2$ between $\left\vert h\right\rangle $ and
$\left\vert v\right\rangle $. If we now apply $U\in SU(2)$, then we obtain $%
UV\left\vert h\right\rangle =\left( e^{-i\phi /2}U\left\vert
h\right\rangle -ie^{i\phi /2}U\left\vert v\right\rangle \right)
/\sqrt{2}$. We are interested in $U_{n}(\theta ,\varphi ,s)=\exp
\left[ -is\mathbf{n}(\theta ,\varphi )\cdot \boldsymbol{\sigma
}/2\right] $ and the geometric phase that this transformation
generates. We recall that the geometric phase is given by
\cite{mukunda}
\begin{equation}
\Phi _{g}(\mathcal{C})=\arg \langle \psi (0)|\psi (s)\rangle
-\Im\int_{0}^{s}\langle \psi
(s^{\prime})|\dot{\psi}(s^{\prime})\rangle ds^{\prime}, \label{1}
\end{equation}%
for a path $\mathcal{C}$ joining the initial state $\left\vert
\psi (0)\right\rangle $ with the final state $\left\vert \psi
(s)\right\rangle $. As already said, $\Phi _{g}$ is invariant
under local gauge transformations.
We exploit this property in order to nullify the dynamical contribution to $%
\Phi _{g}$. That is, we choose a gauge transformation $|\psi
(s)\rangle \rightarrow |\psi ^{\prime }(s)\rangle =\exp \left(
i\alpha (s)\right) |\psi (s)\rangle $ so that $\langle \psi
^{\prime }(s)|\dot{\psi}^{\prime }(s)\rangle =0$. In other words,
instead of applying $U_{n}(\theta ,\varphi ,s)$ we apply $\exp
\left[ i\alpha (s)\right] U_{n}(\theta ,\varphi ,s)$ and measure
the total phase $\arg
\langle \psi(0)|\psi (s)\rangle $. In the present case, this can be achieved by setting $%
\alpha (s)=s\left\langle +\right\vert \mathbf{n}\cdot
\boldsymbol{\sigma }\left\vert +\right\rangle /2$. That
is, we seek to implement the transformation $\left\vert h\right\rangle \rightarrow%
U_{n}V\left\vert h\right\rangle =\left( e^{-i\gamma
/2}U_{n}\left\vert h\right\rangle -ie^{i\gamma /2}U_{n}\left\vert
v\right\rangle \right) /\sqrt{2}$, where $\gamma(s) =\phi -\alpha
(s)$. We can realize this with the help of wave plates. To begin
with, $U_{n}$ can be implemented with a gadget proposed by Simon
and Mukunda \cite{simon}, which is given by
\begin{equation}
U_{n}(\theta ,\varphi ,s)=Q\left( \frac{\pi +\varphi }{2}\right) Q\left(
\frac{\theta +\varphi }{2}\right) H\left( \frac{-\pi +\theta +\varphi }{2}+%
\frac{s}{4}\right) Q\left( \frac{\theta +\varphi }{2}\right) Q\left( \frac{%
\varphi }{2}\right) .  \label{normal}
\end{equation}%
The rotation axis is here given by $\mathbf{n} =(\sin \theta \cos
\varphi ,\sin \theta \sin \varphi ,\cos \theta )$ and the Pauli
matrices are defined according to the convention
that is
commonly employed in optics. That is, the diagonal matrix in the basis $%
\{|h\rangle ,|v\rangle \}$ of horizontally and vertically
polarized states, is $\sigma _{x}$. The other two Pauli matrices
follow from cyclically completing the change $\sigma
_{z}\rightarrow \sigma _{x}$. With this choice, our gauge is given
by
\begin{equation}
\alpha (s)=\frac{s}{2}\sin \theta \cos \varphi .  \label{alpha}
\end{equation}
On the other hand, $V(\gamma )=e^{-i\gamma \sigma _{z}/2} e^{
-i\pi \sigma _{x}/4} $ can be implemented as $V(\gamma) =Q(\pi
/4)H\left(( \gamma -\pi) /4\right)H(\pi /4)$. The total
transformation is thus
\begin{equation}
U_{tot}\equiv V^{\dagger }U_{n}V=H\left( -\frac{\pi }{4}\right) H\left( \frac{%
\gamma +\pi }{4}\right) Q\left( -\frac{\pi }{4}\right)
U_{n}(\theta ,\varphi ,s)Q\left( \frac{\pi }{4}\right) H\left(
\frac{\gamma -\pi }{4}\right) H\left( \frac{\pi }{4}\right) .
\label{up}
\end{equation}
Applying as before relations like $Q(\alpha )H(\beta )=H(\beta
)Q(2\beta -\alpha )$, $Q(\alpha )H(\beta )H(\gamma )=Q(\alpha +\pi
/2)H(\alpha -\beta +\gamma -\pi /2)$, etc., we reduce the above
array to one that consists of seven plates:
\begin{eqnarray}
U_{tot}(\theta ,\varphi ,\phi, s) &=&Q\left( \frac{\pi }{4}-\frac{\gamma_{\phi}(s) }{%
2}\right) Q\left( -\pi -\frac{\varphi
}{2}-\frac{\gamma_{\phi}(s)}{2}\right) Q\left(
\frac{\pi -\theta -\varphi }{2}-\frac{\gamma_{\phi}(s)}{2}\right) \times  \notag \\
&&\times H\left( \frac{-\theta -\varphi
}{2}-\frac{s}{4}-\frac{\gamma_{\phi}(s)}{2}\right) Q\left(
\frac{\pi -\theta -\varphi }{2}-\frac{\gamma_{\phi}(s)}{2}\right)
Q\left(
\frac{\pi -\varphi }{2}-\frac{\gamma_{\phi}(s) }{2}\right) Q\left( -\frac{\pi }{4}-%
\frac{\gamma_{\phi}(s) }{2}\right) ,  \label{2}
\end{eqnarray}%
where $\gamma_{\phi}(s)=\phi -\alpha (s)$. We use this notation to
emphasize that $\gamma$ depends on both $\phi$ and $s$. Note that
by going from Eq.(\ref{up}) to Eq.(\ref{2}) the gauge-fixing role
-- originally played by the plates implementing $V(\gamma)$ --
turns to be shared by all the seven plates of the final array. The
path followed by the polarization state subjected to $U_{tot}$ can
be represented on the Poincar\'{e} sphere by a circular arc, see
Fig.(\ref{path}). This arc is fixed by $\mathbf{n}(\theta ,\varphi
)$, by the initial polarization state, and by $s$. The latter
fixes the angle by which the initial state is rotated. Once we
have fixed $\mathbf{n}$ and the initial state, we record the
geometric phase as a function of $s$. This is done by varying the
registered intensity as a function of $\gamma_{\phi}(s)$, which
plays a double role. First, it contains the phase-shift $\phi $
that is required to implement the polarimetric version of the
Mach-Zehnder interferometer, as discussed above. Second, it
contains the gauge-shift $\alpha(s)$ that is required to make the
total phase coincide with the geometric phase. In order to extract
this geometric phase, we project the state $U_{n}V(\gamma
)\left\vert h\right\rangle $ onto the state $V(\gamma )\left\vert
h\right\rangle =e^{-i\gamma /2}\left( \left\vert h\right\rangle
-ie^{i\gamma }\left\vert
v\right\rangle \right) /\sqrt{2}$. The recorded intensity is thus given by $%
I=\left\vert \left\langle h\right\vert V^{\dagger }(\gamma
)U_{n}V(\gamma )\left\vert h\right\rangle \right\vert ^{2}$. As we
shall see, after having fixed $\theta $, $\varphi $ and $s$, we
can let $\gamma $ (viz. $\phi$) vary so as to generate an
intensity pattern $I(\phi)$, whose maxima and minima determine the
value of the geometric phase at $(s,\theta,\varphi)$. This value can be compared with the theoretical one, which is given by $\Phi _{g}=\Phi _{P}-\Phi _{dyn}$, where
\begin{eqnarray}
\Phi _{P} &=&\arg \langle \psi (0)|\psi (s)\rangle =\arg \left\langle
h\right\vert U_{n}(s)\left\vert h\right\rangle =-\arctan \left[ \sin \theta
\cos \varphi \tan \left( \frac{s}{2}\right) \right] ,  \label{3} \\
\Phi _{dyn} &=&\Im\int_{0}^{s}\langle \psi
(s)|\dot{\psi}(s)\rangle
ds=\Im\int_{0}^{s}\left\langle h\right\vert U_{n}^{\dag }(s)(-i\mathbf{n}%
\cdot \boldsymbol{\sigma })U_{n}(s)\left\vert h\right\rangle ds=-\frac{s%
}{2}\left\langle h\right\vert \mathbf{n}\cdot \boldsymbol{\sigma }%
\left\vert h\right\rangle .  \label{4}
\end{eqnarray}%
The theoretical expression for the geometric phase thus reads
\begin{equation}
\Phi _{g}^{th}=-\arctan \left[ \sin \theta \cos \varphi \tan \left( \frac{s}{%
2}\right) \right] +\frac{s}{2}\sin \theta \cos \varphi .  \label{5}
\end{equation}%
%On the other hand, the intensity reads
%\begin{equation}
%I=\left\vert \left\langle h\right\vert V^{\dagger }\left( \phi
%-\alpha (s)\right) U_{n}(\theta ,\varphi ,s)V\left( \phi -\alpha
%(s)\right) \left\vert h\right\rangle \right\vert ^{2}.  \label{6}
%\end{equation}%
On the other hand, a straightforward calculation of the intensity $I=\left\vert
\left\langle h\right\vert V^{\dagger }\left( \phi -\alpha
(s)\right) U_{n}(\theta ,\varphi ,s)V\left( \phi -\alpha
(s)\right) \left\vert h\right\rangle \right\vert ^{2}$ gives
\begin{equation}
I=\cos ^{2}\left( \frac{s}{2}\right) +\sin ^{2}\left( \frac{s}{2}\right) %
\left[ \cos \theta \cos \left( \alpha(s) -\phi \right) +\sin
\theta \sin \varphi \sin \left( \alpha(s) -\phi \right) \right]
^{2}. \label{7}
\end{equation}%
We have then,
\begin{eqnarray}
I_{\min }(s) &=&\cos ^{2}\left( \frac{s}{2}\right) ,  \label{8a} \\
I_{\max }(s) &=&\cos ^{2}\left( \frac{s}{2}\right) +\sin ^{2}\left( \frac{s}{%
2}\right) \left[ \cos ^{2}\theta +\left( \sin \theta \sin \varphi \right)
^{2}\right] ,  \label{8b}
\end{eqnarray}%
where we have used that the maximum of $f(\alpha )=a\cos \alpha
+b\sin \alpha $ is given by $\sqrt{a^{2}+b^{2}}$. From the above
equations we get
\begin{eqnarray}
\frac{1-I_{\max }}{1-I_{\min }} &=&\sin ^{2}\theta \cos ^{2}\varphi ,
\label{9a} \\
\frac{1-I_{\max }}{I_{\min }} &=&\sin ^{2}\theta \cos ^{2}\varphi \tan
^{2}\left( \frac{s}{2}\right) .  \label{9b}
\end{eqnarray}%
We can thus express $\Phi _{g}^{th}$ in terms of the
experimentally accessible quantities $I_{\min }$ and $I_{\max }$
as
\begin{eqnarray}
\Phi _{g}(s) &=&\sqrt{\frac{1-I_{\max }(s)}{1-I_{\min }(s)}}\arccos \left[
\sqrt{I_{\min }(s)}\right] -\arctan \left[ \sqrt{\frac{1-I_{\max }(s)}{%
I_{\min }(s)}}\right] ,\text{ \ for }-\pi<s<\pi ,  \label{10a} \\
\Phi _{g}(s) &=&\sqrt{\frac{1-I_{\max }(s)}{1-I_{\min }(s)}}\arccos \left[ -%
\sqrt{I_{\min }(s)}\right] +\arctan \left[ \sqrt{\frac{1-I_{\max }(s)}{%
I_{\min }(s)}}\right] \pm \pi ,\text{ \ for }\pi<s<3\pi .
\label{10b}
\end{eqnarray}%
Note that $\Phi _{g}$ is undefined for $s=\pi $, cf. Eq.(\ref{5}). The $%
\pm \pi $ that appears in $\Phi _{g}(s>\pi )$ comes from the
Pancharatnam contribution, $\arg \left\langle h\right\vert
U_{n}(s)\left\vert
h\right\rangle $, that is contained in $\Phi _{g}^{th}$. Indeed, $%
\left\langle h\right\vert U_{n}(s)\left\vert h\right\rangle =\cos
(s/2)[1-i\sin \theta \cos \varphi \tan (s/2)]$, so that $\arg \left\langle
h\right\vert U_{n}(s)\left\vert h\right\rangle =\arg \left( \cos
(s/2)\right) -\arctan [\sin \theta \cos \varphi \tan (s/2)]$. For $\pi
<s<3\pi ,$ we have that $\arg \left( \cos (s/2)\right) =\pm \pi $.

\section{Experimental procedure and analysis of results}

\begin{figure}[tbp]
\begin{centering}
\includegraphics[scale=0.4]{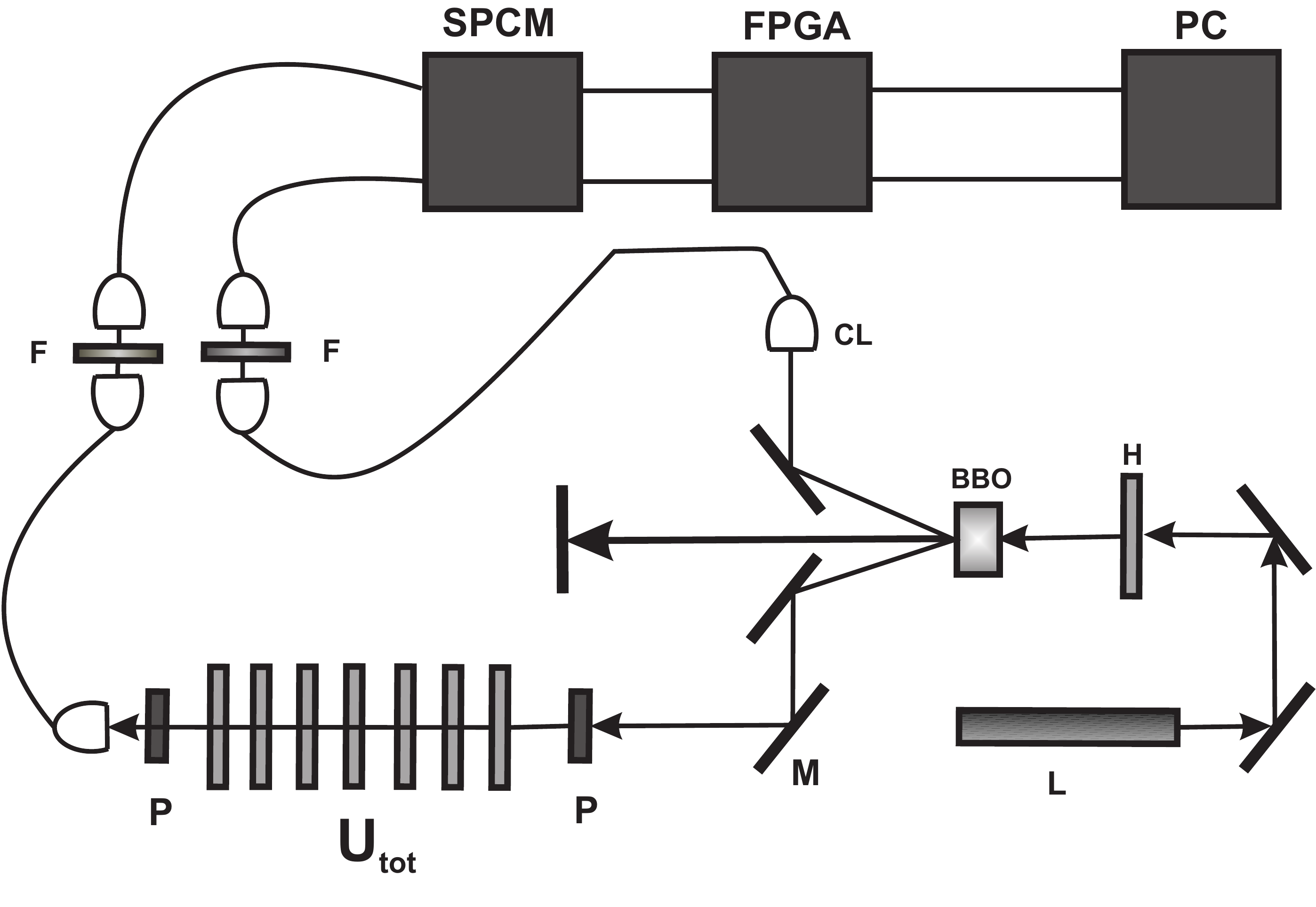} %\includegraphics[angle=-90,scale=1.0]{multi.eps}
\end{centering}
\caption{Polarimetric array. The set of seven wave plates shown at
the bottom can be oriented so as to realize the desired $SU(2)$
transformation ($U_{tot}$) in polarization space. Polarized
photons enter this array after having been produced in a
non-linear, beta-barium borate crystal ($BBO$) that is fed by a
diode laser ($L$) that emits $400$ nm light whose polarization is
fine-tuned with a $\lambda/2$ plate ($H$) placed before the
crystal. Polarizers ($P$) set before and after the retarders
project the photon's polarization as required (see text). Signal
photons are recorded in coincidence with their heralding twins in
a single-photon counting module ($SPCM$). Other components are
$M$: mirrors, $CL$: converging lenses, $F$: filters, $FPGA$: field
programmable gate array, $PC$: personal computer.} \label{array}
\end{figure}

\begin{figure}[tbp]
\begin{centering}
\includegraphics[scale=0.72]{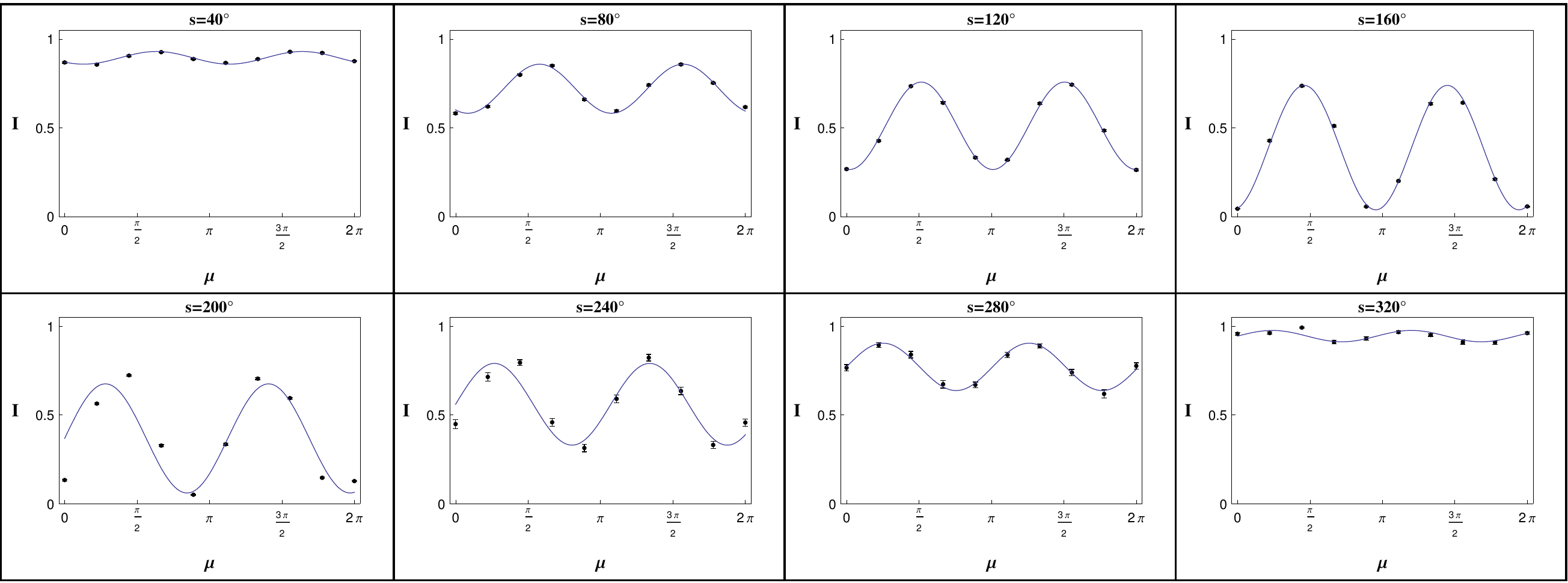} %\includegraphics[angle=-90,scale=1.0]{multi.eps}
\end{centering}
\caption{(Color online) The geometric phase is experimentally
fixed by the maxima and minima of the measured curves
$I_{exp}(\phi)$. The plotted curves correspond to $\theta=\pi/3$,
$\varphi=\pi/3$.} \label{mucurves}
\end{figure}

\begin{figure}[tbp]
\begin{centering}
\includegraphics[scale=0.75]{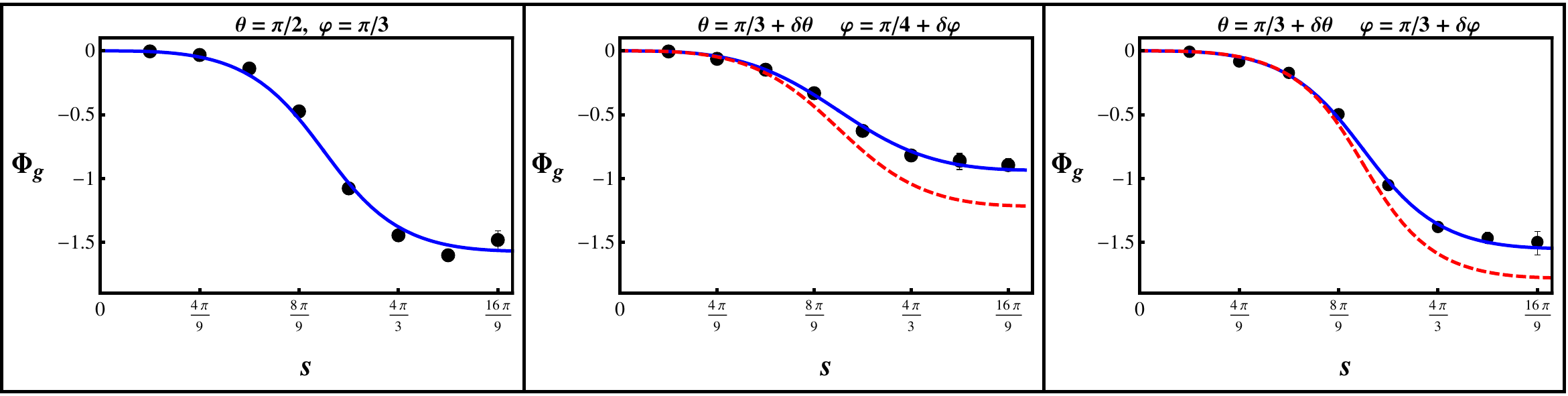} %\includegraphics[angle=-90,scale=1.0]{multi.eps}
\end{centering}
\caption{(Color online) Geometric phase $\Phi
_{g}(s,\theta,\varphi)$ as a function of parameter $s$ for three
choices of $(\theta,\varphi)$. Curve $\Phi _{g}(s,\pi/2,\pi/3)$
closely matches experimental results. However, $\Phi
_{g}(s,\pi/3,\pi/4)$ and $\Phi _{g}(s,\pi/3,\pi/3)$ seem to
systematically deviate from the measured values. By properly identifying the actual values of
$(\theta,\varphi)$, the theoretical curves do match experimental
results. Dashed curves correspond to $\Phi
_{g}(s,\pi/3,\pi/4)$ (middle panel) and to $\Phi
_{g}(s,\pi/3,\pi/3)$ (right panel). Full curves correspond to
$\Phi _{g}(s,\pi/3+\delta\theta,\pi/4+\delta\varphi)$ with $\delta
\theta =3^{\circ}\pi/180^{\circ}$, $\delta \varphi =
-7^{\circ}\pi/180^{\circ}$ (middle panel) and to $\Phi
_{g}(s,\pi/3+\delta\theta,\pi/3+\delta\varphi)$ with $\delta
\theta =5^{\circ}\pi/180^{\circ}$, $\delta \varphi =
-4^{\circ}\pi/180^{\circ}$ (right panel). Most error bars are
smaller than symbols.} \label{geom}
\end{figure}

A sketch of our experimental arrangement is shown in
Fig.(\ref{array}). Its core is the array of seven plates that
realize the transformation $U_{tot}(\theta ,\varphi ,\phi, s)$, as
given in Eq.(\ref{2}). Our single-photon source was a BBO crystal
pumped by a cw diode laser (measured central wavelength: $400$ nm,
spectral line-width lies between $0.5$ and $1$ nm at operating
temperatures; output power: $37.5$ mW). Two photon beams were
produced in the BBO crystal by type-I spontaneous parametric
down-conversion, each beam having a wavelength of $800$ nm. One
beam, the idler or heralding one, was directed towards an
avalanche photodetector. The other, signal beam, was directed
towards the array of seven plates. Coincidence counts ($I$) of
idler and signal beams made up our raw data, with coincidences
being defined within a time-window of $10.42$ ns. Our
photon-counting module was a Perkin-Elmer SPCM-AQ4C, with a dark
count-rate of $500\pm10$ cps. Photons were collected with the help
of converging lenses that focused them into multimode fiber optic
cables having fiber-coupling connectors at both ends. The recorded
coincidences were obtained according to the following procedure.
For given values of $\theta $, $\varphi $ and $s$, the seven
plates were oriented as prescribed in Eq.(\ref{2}), with $\gamma
=\phi -s\sin \theta \cos \varphi /2$. The angle $\phi $ was varied
from $0^{\circ}$ to $360^{\circ}$ in steps of $40^{\circ}$.
Coincidence counts were recorded as a function of $\phi $ and then
normalized to obtain the intensity $I(\phi)$. Theoretically,
$I(\phi)$ is given by Eq.(\ref{7}), with $s$, $\theta$ and
$\varphi$ being kept fixed. By repeated measurements we sampled
$30$ points for each value of $\phi $. The parameter $s$ took
values $s_{i}$ from $40^{\circ}$ to $320^{\circ}$ in steps of
$40^{\circ}$. After averaging the recorded coincidence counts for
each $\phi $ we obtained a series of points $I(\phi_{i})$. A best
fit $I(\phi)$ to these points was found, where $I(\phi)$ is a
sinusoidal function whose parameters were fixed by the least
squares method. Fig.(\ref{mucurves}) shows the so obtained curves
for $\theta=\pi/2$, $\varphi=\pi/3$ and different values of $s$.
From these curves we determined $I_{\max }$ and $I_{\min}$.
Entering $I_{\max }$ and $%
I_{\min }$ in Eqs. (\ref{10a}) and (\ref{10b}), the experimental values of $%
\Phi _{g}(s,\theta,\varphi)$ can be obtained and compared with the ones predicted by Eq.(%
\ref{5}). Fig.(\ref{geom}) shows our experimental results together
with the corresponding theoretical predictions. As can be seen,
two of the three cases seem to reflect a systematic departure of
our experimental findings from the theoretical predictions. We
will come back to this point below. As for the single-photon
production, it was checked by the standard procedure
\cite{kwiat,thorn} of measuring the degree of second-order
coherence, $g^{(2)}$, between the output fields of a
beam-splitter, i.e., the reflected (R) and transmitted (T) beams.
Detections at gates T and R were
conditioned upon detection at a third gate G. In such a case, $%
g^{(2)}=P_{GTR}/(P_{GT}P_{GR})$, where the $P_{a}$ denote
probabilities for simultaneous detection at gates specified by
label $a$. In terms of photocounts,
$N_{a}$, the degree of coherence can be expressed as \cite{grangier} $%
g^{(2)}=N_{GTR}N_{G}/(N_{GT}N_{GR})$. It has a value that is less
than $1$ for non-classical light. We obtained $g^{(2)}= 0.187 \pm
0.011$ in our experiments.

Several sources of experimental error could be identified. The
main source of error came from the accuracy with which our plates
could be oriented, i.e., $\pm 1^{\circ}$ approximately. Another
possible source of error came from our photons having a wavelength
of $800$ nm instead of the $808$ nm that would be required for
optimal performance of our wave plates. These are zero-order
plates whose effective retardances at the produced wavelength made
them slightly differ from being $\lambda /2$ and $\lambda /4$
plates.
However, the corresponding departures ($0.505\lambda $ instead of $\lambda /2$ and $%
0.253\lambda $ instead of $\lambda /4$) were small enough to be
neglected as a sensible source of error. Accidental coincidence
counts were also estimated to be too small (contribution to
$g^{(2)}$ less than $0.19$) for them to have a noticeable
influence on the departures of our experimental findings from the
theoretically predicted values when $s>\pi $ (see
Fig.(\ref{geom}), middle and right panels). As illustrated in
Fig.(\ref{geom}), left panel, the agreement between theoretical
predictions and measured values was very good. However, we also
observed slight departures that occasionally increased. The dashed
curves in Fig.(\ref{geom}), middle and right panels, correspond to
the targeted geometric phase $\Phi_{g}(s,\theta,\varphi)$. Large
departures seemed to reflect a drift of the measured values with
respect to the assumed theoretical curve, rather than random
fluctuations around this curve. In what follows, we substantiate
our claim that the $\pm 1^{\circ}$ accuracy in the orientation of
our plates does explain occasional, systematic departures of
experimental measurements from theoretical predictions. Depending
on the measured quantity, rotation errors of this magnitude can
give rise to inaccuracies of various sorts, like those recently
reported in \cite{fisher}. It is important to identify error
sources and their effects, specially when one's ultimate goal is
to have a good understanding of how the geometric phase behaves in
a noisy environment.

Let us denote by $\delta _{i}$ the departure of the $i$-th plate's
orientation from its nominal value. For a quarter-wave plate we must then
set $Q(x+\delta )$ instead of $Q(x)$ in Eq.(\ref{2}). To first order in $%
\delta $, we get $dQ(x)=Q(x+\delta )-Q(x)=\sqrt{2}i\delta R_{x}$,
with
\begin{equation}
R_{x}=\left(
\begin{array}{cc}
\sin (2x) & -\cos (2x) \\
-\cos (2x) & -\sin (2x)%
\end{array}%
\right) .  \label{11}
\end{equation}%
Similarly, for a half-wave plate we obtain $dH(x)=H(x+\delta
)-H(x)=2i\delta R_{x}$. If we now replace the operators $Q(x)$ and
$H(x)$ in Eq.(\ref{2}) by $Q(x)+dQ(x)$ and $H(x)+dH(x)$,
respectively, and then expand the result to first order in the
$\delta_{i} $, we obtain
\begin{equation}
U_{tot}^{\delta }=U_{tot}+\sum_{i=1}^{7}U_{i}^{\delta },  \label{12}
\end{equation}%
where $U_{i}^{\delta }$ reads like $U_{tot}$, see Eq.(\ref{2}),
except that its $i$-th factor is replaced by $dH(x)$ when $i=4$
and by $dQ(x)$ otherwise. $U_{tot}^{\delta }$ is then a function
of all $\delta _{i=1,\ldots ,7}$. From the amplitude $\left\langle
h\right\vert U_{tot}^{\delta }\left\vert h\right\rangle $, we can
calculate the total intensity $I_{\delta }=\left\vert \left\langle
h\right\vert U_{tot}^{\delta }\left\vert h\right\rangle
\right\vert ^{2}$, once again to first order in the $\delta _{i}$.
With this expression, by choosing different values for the $\delta
_{i}$, we can study how much $I_{\delta }(\phi )$ differs from the
$I(\phi )$ given in Eq.(\ref{7}). We have found that the
departures from $I$ can be very sensitive to a change from, say,
$\delta _{i}\approx +1^{\circ}$ to $\delta _{i}\approx
-1^{\circ}$, keeping fixed all the other $\delta _{j\neq i}$. The
values of $I_{\max }$ and $I_{\min }$ can be calculated using
$I(\phi )$ and $I_{\delta }(\phi)$, in order to assess the
sensitivity of the array to changes $\delta _{i}\approx \pm
1^{\circ}$ in the setting of the plates. The values of $I_{\max }$
and $I_{\min }$ that correspond to $I_{\delta }(\phi)$ show that
inaccuracies $\delta _{i}\approx \pm 1^{\circ}$ can explain the
observed
differences between recorded phases and theoretically predicted ones, cf. Eqs. (\ref{10a}) and (\ref%
{10b}).

Last claim can be confirmed by the following, independent
approach. Inaccuracies $\delta _{i}\approx \pm 1^{\circ}$ should
translate into a departure of $\theta $ and $\varphi $ from their
nominal values. Let us then assume that our array does not realize
the transformation $U_{n}(\theta ,\varphi ,s)=\exp \left[
-is\mathbf{n}(\theta ,\varphi )\cdot
\boldsymbol{\sigma }/2\right] $, but instead $\exp \left[ -is\mathbf{n}%
(\theta +\delta \theta ,\varphi +\delta \varphi )\cdot \boldsymbol{%
\sigma }/2\right] $, with $\delta \theta \approx \pm
7^{\circ}\approx
\delta \varphi $. The actual values of $%
\delta \theta $ and $\delta \varphi $ can be obtained by the
following procedure. From Eq.(\ref{9a}) we see that $I_{\max
}(s_{i})$ and $I_{\min }(s_{i})$ corresponding to targeted values
$\theta $ and $\varphi $ should satisfy
\begin{equation}
y(s_{i})\equiv \frac{1-I_{\max }(s_{i})}{1-I_{\min }(s_{i})}=\sin ^{2}\theta
\cos ^{2}\varphi \equiv f(\theta ,\varphi ).  \label{13}
\end{equation}%
The above equation can be used to determine the actual values of $\theta $
and $\varphi $, i.e., $\theta +\delta \theta $ and $\varphi +\delta \varphi $%
, by the least squares method. To this end, we evaluate the
right-hand side of Eq.(\ref{13}) in the sought-after values,
expand it to first order, i.e., we set $f(\theta +\delta \theta
,\varphi +\delta \varphi )=f(\theta ,\varphi )+\left( \sin2 \theta
\cos^{2} \varphi \right) \delta \theta -\left( \sin^{2} \theta
\sin 2\varphi \right) \delta \varphi $, and then determine $\delta
\theta $, $\delta \varphi $ as
\begin{equation}
\left(
\begin{array}{c}
\delta \theta \\
\delta \varphi%
\end{array}%
\right) =\left( A^{T}WA\right) ^{-1}A^{T}Wb.  \label{14}
\end{equation}%
Here, $(\cdot)^{-1}$ means the Moore-Penrose pseudoinverse, $b$ is the column vector $(y(s_{i})-f(\theta ,\varphi ))^{T}$, with $%
i=1,\ldots ,n$\ ($n$ being the number of recorded points), $A$ is the $%
n\times 2$ matrix whose rows are all equal to $(\sin 2\theta%
\cos^{2} \varphi , -\sin^{2}\theta \sin 2\varphi )$ and $W$ is the
inverse of the covariance matrix, i.e., $W=\diag(\sigma
_{1}^{-2},\ldots ,\sigma _{n}^{-2})$. The latter corresponds to
statistically uncorrelated measurements having different variances
$\sigma_{i}$ at different values $s_{i}$. We have assessed these
variances in two different ways. First, by fitting a Gaussian to
the distributions of measured points, cf. Fig.(\ref{mucurves}),
which gives us $\sigma_{i}$ for each value $I(\phi_{j})$ and hence
for $I_{\min}$, $I_{\max}$, and $\Phi_{g}$ by error propagation.
Second, from our raw data, which consists of $30$ values for each
$\phi_{i}$ -- with $s$, $\theta$, $\varphi$ being kept fixed -- we
randomly chose $10$ values for each $\phi_{i}$ and calculated
$\Phi_{g}$ as we did when using the $30$ values. By iterating this
procedure several times ($\approx 40$), we got a series of values
for each $\Phi_{g}(s,\theta,\varphi)$. From each series we
obtained a mean value and its corresponding maximal and minimal
departures. These departures constitute our error bars. Such an
estimation is justified by the statistical independence of our
measurements. Thus, randomly sampling $10$ out of $30$ measured
values amounts to having recorded $10$ values in each run of the
experiment while repeating it several times ($\approx 40$). From
the two methods we observe that our measured values $\sigma_{i}$
span a range that goes from a minimum of $1.3\times10^{-4}$ to a
maximum of $0.12$. The plotted error bars, cf. Fig(\ref{geom}),
are mostly smaller than the symbols and can be barely seen only in
cases for which $\sigma_{i}\approx 0.1$. Now, the above mentioned
application of the least squares method holds whenever
inaccuracies $\delta s_{i}$ of the $s_{i}$ can be neglected. In
our case, the nominal value of $s$ enters in the orientations of
our wave plates, and the inaccuracies of these orientations are
precisely the assumed main source of errors. Nevertheless, the
above application of the least squares method is justified.
Indeed, we can assess the values of the $\delta  s_{i}$ by using
Eq.(\ref{8a}). That is, we set $\delta s_{i}\approx
\left|s_{i}-2\arccos\left(\sqrt{I_{\min}(s_{i})}\right)\right|$ as
an estimator of the inaccuracies of the $s_{i}$. These
inaccuracies turn out to be negligible in comparison to our
$\sigma_{i}$ -- besides, if they were not, they would modify the
above results only to higher order than the first in
$(\delta\theta, \delta\varphi)$, because our $y(s_{i})$ do not
depend on $s$, as Eq.(\ref{13}) shows. The least squares method
can thus be iteratively applied to find successive values of
$\delta \theta$ and $\delta \varphi$, until $\Phi _{g}^{th}(s,
\theta +\delta \theta,\varphi +\delta \varphi )$ of Eq.(\ref{5})
eventually matches experimental results. In the present case,
however, it proved more practical to seek the right choice of
$\theta $ and $\varphi $ by hand, i.e., by trial and error when
plotting $\Phi _{g}^{th}(s,\theta +\delta \theta ,\varphi +\delta
\varphi )$ together with its measured values. Indeed, by so doing
in the cases of Fig.(\ref{geom}), middle and right panels, we
quickly found values $\delta \theta \approx \pm 7^{\circ}\approx
\delta \varphi $ for which the theoretical curves approximate very
closely our experimental results. Fig.(\ref{geom}) shows the
curves obtained with $\delta \theta =3^{\circ}$, $\delta \varphi =
-7^{\circ}$ (middle panel) and $\delta \theta =5^{\circ}$, $\delta
\varphi = -4^{\circ}$ (right panel). Such a result is consistent
with the assumed errors $\delta _{i}\approx \pm 1^{\circ}$, which
may accumulate so as to produce inaccuracies $\delta \theta
\approx \pm 7^{\circ}\approx \delta \varphi $. Thus, departures of
$\theta $ and $\varphi $ from their targeted values do explain our
experimental findings. We have thereby assessed the amount by
which the theoretically predicted value $\Phi
_{g}^{th}(s,\theta,\varphi)$ might differ from the experimentally
realized one. Such a difference should be taken into account when
assessing with the help of a polarimetric array the robustness of
$\Phi _{g}$ against decohering mechanisms.

Finally, let us point out the following feature of our array. As
can be seen from Eqs.(\ref{10a}) and (\ref{10b}), the geometric
phase we produce depends on $\theta$ and $\varphi$ only through
$|\sin\theta\cos\varphi|$. This means that we can fix the actually
realized values of $\theta$ and $\varphi$ only up to changes
$(\theta,\varphi)\rightarrow(\theta^{\prime},\varphi^{\prime})$
that leave $|\sin\theta\cos\varphi|$ invariant. Instead of seing
this as a weakness of our approach, such a feature can be helpful
when seeking to exploit the robustness of $\Phi_{g}$ against
decoherence. Indeed, if one is able to confine decohering effects
to those regions in the plane $(\theta,\varphi)$ for which the
variations in $|\sin\theta\cos\varphi|$ are sufficiently small,
then $\Phi_{g}$ will vary also within acceptable limits. Of
course, these limits will depend on the application one has in
mind and on the decohering mechanisms, which should be studied in
detail. Such an endeavor goes beyond the scope of the present
paper and is deferred to future work.

\section{Conclusions}

Our polarimetric setup proved to be a versatile tool for testing
geometric phases. The main part of it, an array made of one
$\lambda/2$ and six $\lambda/4$ plates, allows us to realize
geometric phases that are associated to non-geodesic paths on the
Poincar\'{e} sphere. Although we have limited ourselves to study
circular trajectories, our approach can be extended to deal with
arbitrary paths. Our experimental results fit very closely
theoretical predictions, once we have accurately identified the
trajectory on the Poincar\'{e} sphere that has been actually
realized by our setting. The end product of such a setting is a
geometric phase $\Phi_{g}$ that is non-trivially related to
various parameters entering our setup. Indeed, coincidence counts
must be optimized by adjusting the laser polarization, the
acquisition window for photon counts must also be properly fixed,
and the wave plates must be repeatedly set to their nominal
orientations when recording the data from which $\Phi_{g}$ can be
extracted. Not only because of the photon counting statistics, but
mainly because of our $\pm 1^{\circ}$ accuracy in the setting of
the plates, one could expect experimental results falling within
some region around the theoretical curves, as reported, e.g., in
\cite{fisher}. If that would have been the case, our polarimetric
array would have proved inappropriate for studying robustness of
geometric phases against noise. However, our array does produce
geometric phases that are in accordance with theoretical
expressions. Occasionally, these expressions must be evaluated
\textit{a posteriori}, thereby identifying the actually realized
values of the parameters fixing $\Phi_{g}$. Once the value of
$\Phi_{g}$ has been fixed, our array could be used for assessing
the robustness of this $\Phi_{g}$ against noise. To this end, the
array must be complemented so as to simulate different kinds of
noise. For instance, one can replace the single-crystal photon's
source and use instead polarization-entangled photons produced by
parametric down-conversion in a two-crystal geometry
\cite{kwiat2,walborn}. This produces variable entangled
polarization states. After tracing over the polarization of one of
these photons, its twin photon is brought into a mixed
polarization state $\rho=\left( \openone +r\mathbf{n}\cdot
\boldsymbol{\sigma }\right)/2$, with $r\in[0,1]$ being the degree
of polarization. Such a state can be submitted to a polarimetric
array similar to the one discussed in this paper. Now, $\rho$ can
be written in the form $\rho=\lambda_{+}\left\vert
\mathbf{n}_{+}\right\rangle \left\langle \mathbf{n}_{+}\right\vert
+\lambda_{-}\left\vert \mathbf{n}_{-}\right\rangle \left\langle \mathbf{n}%
_{-}\right\vert$, with $\lambda _{\pm }=\left( 1\pm r\right) /2$
and $\mathbf{n}\cdot \boldsymbol{\sigma }\left\vert
\mathbf{n}_{\pm }\right\rangle =\pm \left\vert \mathbf{n}_{\pm
}\right\rangle$. Applying to $\left\vert \mathbf{n}_{\pm
}\right\rangle$ the techniques of the present work one can get the
corresponding (pure-state) geometric phases $\pm\Phi_{g}$. This is
all one needs \cite{larssons} to obtain the geometric phase of the
mixed-state $\rho$, thereby assessing the effect of noise.
Experiments along these lines have been already performed in
neutron polarimetry \cite{klepp2,filipp}. The kind of noise
studied in \cite{klepp2} translated into a Stokes vector
$\mathbf{r}=r\mathbf{n}$ of the restricted form
$\mathbf{r}=(0,-r,0)$, and the explored paths on the Bloch sphere
originated from unitary transformations that depended on two of
the three Euler angles \cite{klepp2}. By appropriate choice of
these two angles one can generate purely geometric, purely
dynamical, or combinations of both phases. However, once this
choice is made, one cannot freely address different paths on the
Bloch sphere. Nevertheless, these results represented a
considerable extension of previous ones \cite{klepp}, which dealt
with Pancharatnam's phase only. Further progress in assessing the
robustness of geometric phases was achieved by addressing
adiabatic evolutions \cite{filipp}. Here, the dynamical
contribution to the total phase was eliminated by spin-echo
techniques, which impose some restrictions on the class of paths
being explored. Our all-optical setting offers some advantages as
compared to neutron polarimetry. It allows choosing arbitrary
paths on the Poincar\'{e} sphere, as well as different kinds of
noise to be explored in conjunction with the chosen path. The
aforementioned remote state preparation of mixed states is not the
only choice. One can also employ interferometric techniques to
produce an enlarged family of mixed states \cite{fisher,fdz2}. By
applying interferometry for input-state preparation and
polarimetry for state manipulation, one has the possibility of
studying the resilience of purely geometric phases to various
types of noise.

\section{Acknowledgements}
This work was partially financed by DGI-PUCP (Grant No.
2013-0130). J. C. S. thanks E. J. Galvez for support and advice
during his stay at Colgate University.

\newpage

%\begin{figure}[tbp]
%\begin{centering}
%\end{centering}
%\caption{(Color online) Geometric phase $\Phi
%_{g}(s,\theta,\varphi)$ as a function of parameter $s$ for three
%choices of $(\theta,\varphi)$. The curve that corresponds to $\Phi
%_{g}(s,\theta,\varphi)$ evaluated at the nominal values
%$(\theta=\pi/2,\varphi=\pi/3)$ closely matches experimental
%results, see left panel. However, the curves
%on the middle and right panels seem to systematically deviate from
%the measured values. Fig.(\ref{geom}) shows that by properly
%identifying the actually realized values of $(\theta,\varphi)$,
%the theoretical curves do match experimental values.}
%\label{geom3}
%\end{figure}

%\begin{figure}[tbp]
%\begin{centering}
%\includegraphics[scale=0.8]{fig4.eps} %\includegraphics[angle=-90,scale=1.0]{multi.eps}
%\end{centering}
%\caption{(Color online) Theoretically predicted coincidences $I$
%as a function of $\phi$, for different values of $s$ and $\theta=
%\varphi=\pi/3$. In each plot, one curve corresponds to a fully
%accurate setting of the wave plates in the polarimetric array,
%while the other curve takes into account possible deviations from
%the accurate setting. The geometric phase is experimentally fixed
%by the maxima and minima of the measured curves $I_{exp}(\phi)$.}
%\label{icurves}
%\end{figure}

\begin{thebibliography}{99}
\bibitem{berry} M. V. Berry, Proc. R. Soc. London Ser. A \textbf{392}, 45 (1984).

\bibitem{hannay} J. H. Hannay, J. Phys. A \textbf{18}, 221 (1985).

\bibitem{tong} D. M. Tong, E. Sj\"{o}qvist, L. C. Kwek, and C. H. Oh, Phys. Rev. Lett. \textbf{93}, 080405 (2004).

\bibitem{peixoto} J. G. Peixoto de Faria, A. F. R. de Toledo Piza, and M. C. Neves,
Europhys. Lett. \textbf{62}, 782 (2002).

\bibitem{cucchietti} F. M. Cucchietti, J.-F. Zhang, F. C. Lombardo, P. I. Villar, and R. Laflamme,
Phys. Rev. Lett. \textbf{105}, 240406 (2010).

\bibitem{uhlmann} A. Uhlmann, Rep. Math. Phys. \textbf{24}, 229 (1986).

\bibitem{sjoqvist} E. Sj\"{o}qvist, A. K. Pati, A Ekert, J. S. Anandan, M. Ericsson, D. K. L. Oi, and V. Vedral,
Phys. Rev. Lett. \textbf{85}, 2845 (2000).

\bibitem{marzlin} K.-P. Marzlin, S. Ghose, and B. C. Sanders, Phys. Rev. Lett. \textbf{93},
260402 (2004).

\bibitem{bassi} A. Bassi and E. Ippoliti, Phys. Rev. A
\textbf{73}, 062104 (2006).

\bibitem{rezakhani} A. T. Rezakhani and P. Zanardi, Phys. Rev. A
\textbf{73}, 052117 (2006).

\bibitem{yin} S. Yin and D. M. Tong, Phys. Rev. A
\textbf{79}, 044303 (2009).

\bibitem{buric} N. Buri\'{c} and M. Radonji\'{c}, Phys. Rev. A
\textbf{80}, 014101 (2009).

\bibitem{lombardo} F. Lombardo and P. I. Villar, Phys. Rev. A \textbf{81}, 022115 (2010).

\bibitem{sjoqvist2} E. Sj\"{o}qvist,
Physics \textbf{1}, 35 (2008).

\bibitem{klepp} J. Klepp, S. Sponar, Y. Hasegawa, E. Jericha, G. Badurek,
Phys. Lett. A \textbf{342}, 48 (2005).

\bibitem{bertlmann} R. Bertlmann, K. Durstberger, Y. Hasegawa, and
B. C. Hiesmayr, Phys. Rev. A \textbf{69}, 032112 (2004).

\bibitem{klepp2} J. Klepp, S. Sponar, S. Filipp, M. Lettner, G.
Badurek, and Y. Hasegawa, Phys. Rev. Lett. \textbf{101}, 150404
(2008).

\bibitem{filipp} S. Filipp, J. Klepp, Y. Hasegawa, C. Plonka-Spehr, U. Schmidt, P. Geltenbort, and H. Rauch,
 Phys. Rev. Lett. \textbf{102}, 030404 (2009).

\bibitem{werner} S. Werner, Found. Phys. \textbf{42}, 122 (2012).

\bibitem{wagh1} A. G. Wagh and V. C. Rakhecha,
Phys. Lett. A \textbf{197}, 112 (1995).

\bibitem{sponar} S. Sponar \textit{et al.}, New J. Phys.
\textbf{14}, 053032 (2012).

\bibitem{mukunda} N. Mukunda and R. Simon,
Ann. Phys. (N. Y.) \textbf{228}, 205 (1993).

\bibitem{pancharatnam} S. Pancharatnam,
Proceedings of the Indian Academy of Science A \textbf{44}, 247
(1956).

\bibitem{Audretsch} J. Audretsch, \textit{Entangled systems: new directions in quantum physics
}, Wiley-VCH, Chichester (2007).

\bibitem{loredo} J. C. Loredo, O. Ort\'{i}z, R. Weing\"{a}rtner, and F. De Zela,
Phys. Rev. A \textbf{80}, 012113 (2009).

\bibitem{fdz} J. C. Loredo, O. Ort\'{i}z, A. Ball\'{o}n, and F. De Zela,
J. of Phys.: Conf. Ser. \textbf{274}, 012140 (2011).

\bibitem{loredo2} J. C. Loredo, Master Thesis, Physics
Section, Pontificia Universidad Cat\'{o}lica del Per\'{u}, Lima-Peru (2011).

\bibitem{simon} R. Simon and N. Mukunda,
Phys. Lett. A \textbf{138}, 474 (1989).

\bibitem{kwiat} P. G. Kwiat and R. Y. Chiao
Phys. Rev. Lett. \textbf{66}, 588 (1991).

\bibitem{thorn} J. J. Thorn, M. S. Neel, V. W. Donato, G. S. Bergreen, R. E. Davies, and M. Beck,
Am. J. Phys. \textbf{72}, 1210 (2004).

\bibitem{grangier} P. Grangier, G. Roger, and A. Aspect, Europhys.
Lett. \textbf{1}, 173 (1986).

\bibitem{kwiat2} P. Kwiat, E. Waks, A. G. White, I. Appelbaum, and
P. H. Eberhard, Phys. Rev. A \textbf{60}, R773 (1999).

\bibitem{walborn} S. P. Walborn, C. H. Monken, S. P\'{a}dua, and P. H. Souto Ribeiro, Phys. Rep. \textbf{495},
87 (2010).

\bibitem{larssons} P. Larsson and E. Sj\"{o}qvist, Phys. Lett. A
\textbf{315}, 12 (2003).

\bibitem{fisher} K. A. G. Fisher, R. Prevedel, R. Kaltenbaek, and
K. J. Resch, New J. Phys. \textbf{14}, 033016 (2012).

\bibitem{fdz2} F. De Zela, J. Opt. Soc. Am. A \textbf{30} 1544 (2013).



\end{thebibliography}
\end{document}